\documentclass[pra,showpacs,preprintnumbers,amsmath,amssymb]{revtex4}
\usepackage{amsmath}

\usepackage{graphicx}
\usepackage{dcolumn}
\usepackage{bm}
\usepackage{mathrsfs}
\usepackage{graphicx}
\usepackage{float}

\begin{document}

\title{ Generation of four-photon polarization entangled state based on EPR entanglers }

\author{Meiyu Wang, Fengli Yan}

\email{flyan@mail.hebtu.edu.cn}

\affiliation {
    College of Physics Science and Information Engineering, Hebei Normal University, Shijiazhuang 050016, China}
\date{\today}

\begin{abstract}
{We show how to prepare four-photon polarization entangled states
based on some  Einstein-Podolsky-Rosen (EPR) entanglers. An EPR entangler consists of two
single photons, linear optics elements, quantum non-demolition
measurement using a weak cross-Kerr nonlinearity,  and classical
feed forward. This entangler which acts as the most primary part in
the construction of our scheme allows us to make two separable
polarization qubits entangled near deterministically. Therefore, the
efficiency of the present device  completely depends
on that of  EPR entanglers, and it has a high success probability.}
\end{abstract}
 \maketitle

\section{Introduction}
 Entanglement plays an important role in quantum information processing (QIP).
  Although most of the research in QIP are concerned with bipartite
systems, multipartite entanglement has also attracted increasing
interest since it is more stable and has a longer decoherence time \cite {Briegel, Dur, Scarani,Raussendorf,Zhou,Chen, Dong, Zheng, Sun, Brown, Yan, Gao, GaoHong, HongGaoYan, DingYan}.
In Ref.\cite {Briegel}, Briegel et al.  introduced a special type of
multipartite entangled states, i.e., the so-called cluster states, which
 have some special features. Cluster states share the properties
of both Greenberger-Horne-Zeilinger (GHZ) and W class entangled
states. While they have a high persistence of entanglement and can
be regarded as an entanglement source for the GHZ state \cite {Briegel}, but are
more immune to decoherence than GHZ states \cite {Dur}. It has been shown
that a new inequality is maximally violated by the four-particle
cluster states but not by the four-particle GHZ states, and the
cluster states can also be used to test nonlocality without
inequalities \cite {Scarani}. In addition, the cluster states have been
recognized as the basic building blocks for one-way quantum
computation proceeding only by local measurements and feed forward
of their outcomes \cite {Raussendorf}. Since its particular properties, the cluster
states have attracted  much more attention in theoretical research
and practical applications \cite {Zhou,Chen, Dong, Zheng,Sun,Brown}.

 Recently, experiments with
entangled photons open a broad field of research. Several schemes
have been proposed for generation of polarization entangled cluster
states. In particular, Walther et al. \cite {Walt}  experimentally
generated four-photon cluster states and demonstrated the
feasibility of the one-way quantum computation. Zou and
Mathis \cite {Zou}  proposed an experimentally feasible scheme for
preparing a four-photon polarization entangled cluster state of the
form
\begin{equation}\label{1}
 |cluster\rangle=\frac{1}{2}(|H\rangle_{1}|H\rangle_{2}|H\rangle_{3}|H\rangle_{4}+|H\rangle_{1}|H\rangle_{2}|V\rangle_{3}|V\rangle_{4}
                   +|V\rangle_{1}|V\rangle_{2}|H\rangle_{3}|H\rangle_{4}-|V\rangle_{1}|V\rangle_{2}|V\rangle_{3}|V\rangle_{4}),
\end{equation}
where $H$ and $V$ denote horizontal and vertical linear polarizations
respectively, and subscripts $i$ ($i=1,2,3,4$) denote the spatial
modes of photons.

  Another  example of multipartite entangled states is a
four-qubit entangled state of the polarization form
\begin{equation}\label{2}
 |\chi\rangle=\frac{1}{2}[(|H\rangle_{1}|H\rangle_{2}+|V\rangle_{1}|V\rangle_{2})|H\rangle_{3}|H\rangle_{4}
                 +(|H\rangle_{1}|V\rangle_{2}+|V\rangle_{1}|H\rangle_{2})|V\rangle_{3}V\rangle_{4}],
\end{equation}
whose importance has been pointed out by Gottesman and Chuang \cite {Gottesman} that
it can be used as a resource for teleportation of two qubits. The state $|\chi\rangle$ is also important in
the sense that it is equivalent to the cluster state described by
Eq.(1) under a local unitary transformation. Several schemes \cite {Hofmann,Pittman,Tokunaga,Wang,Zhu}
can be utilized to prepare $|\chi\rangle$.

 In this paper, we propose a simple experimental scheme for the preparation of the
 four-photon polarization entangled states $|cluster\rangle$ and
 $|\chi\rangle$ based on EPR entanglers described in Ref.\cite {Nemoto}, which involves some basic elements: the polarizing beam
 splitters (PBSs), the cross-Kerr nonlinearity and quantum non-demolition
 detection (QND) with feed forward. In Sec. II, we first review the cross-Kerr nonlinear
interaction, then describe how the EPR entangler works. In Sec.
III, we show an experimental scheme  for generation of four-photon
polarization entangled states. Finally, we describe our conclusions
in Sec. IV.

\section{cross-Kerr nonlinearity and near deterministic EPR entangler }
  In the process of preparing the entangled states in EPR entanglers, one of the basic
elements is cross-Kerr nonlinearity, which was first used by Chuang
and Yamamoto to realize the simple optical quantum computation \cite {Chuang}.
Let us briefly review the cross-Kerr nonlinearity interaction
between a signal mode and a probe mode. The interaction Hamiltonian
has the form
$\hat{H}_{k}=-\hbar\kappa\hat{a}^{\dag}_{s}\hat{a}_{s}\hat{a}^{\dag}_{p}\hat{a}_{p}$,
where the signal (probe) mode has the creation and annihilation
operators given by $\hat{a}^{\dag}_{s},\hat{a}_{s}$
($\hat{a}^{\dag}_{p},\hat{a}_{p}$) respectively, and $\kappa$ is the
strength of the nonlinearity. If we consider the signal state to
have the form $|\psi\rangle=c_{0}|0\rangle_{s}+c_{1}|1\rangle_{s}$
with the probe beam initially in a coherent state
$|\alpha\rangle_{p}$, the cross-Kerr nonlinearity interaction causes
the combined signal-probe system to evolve as
\begin{equation}\label{3}
\mathrm{e}^{-\mathrm{i}\hat{H}_{k}t/\hbar}|\psi\rangle_{s}|\alpha\rangle_{p}=c_{0}|0\rangle_{s}|\alpha\rangle_{p}
+c_{1}|1\rangle_{s}|\alpha \mathrm{e}^{\mathrm{i}\theta}\rangle_{p},
\end{equation}
where $\theta=\kappa t$ with $t$ being the interaction time. It is
easy to observe that the Fock state is unaffected by the interaction
but the coherent state picks up a phase shift directly proportional
to the number of photons $n_{s}$ in the signal mode. One can exactly
obtain the information of photons in the Fock state but not destroy
them by detecting the probe mode. We called this an X homodyne
measurement. In the following, we will illustrate the EPR entangler
with the help of the cross-Kerr effect. In Ref.\cite {Nemoto}, Nemoto and
Munro  constructed a two qubit polarization parity QND detector,
which can be configured to act as a near deterministic entangler.
Therefore,  here and hereafter we call it the EPR entangler for
simplicity. This gate allows us to make two separable polarization
qubits entangled efficiently (near deterministically). The
construction of the EPR entangler has been shown in Fig.1.

\begin {center}
\includegraphics* {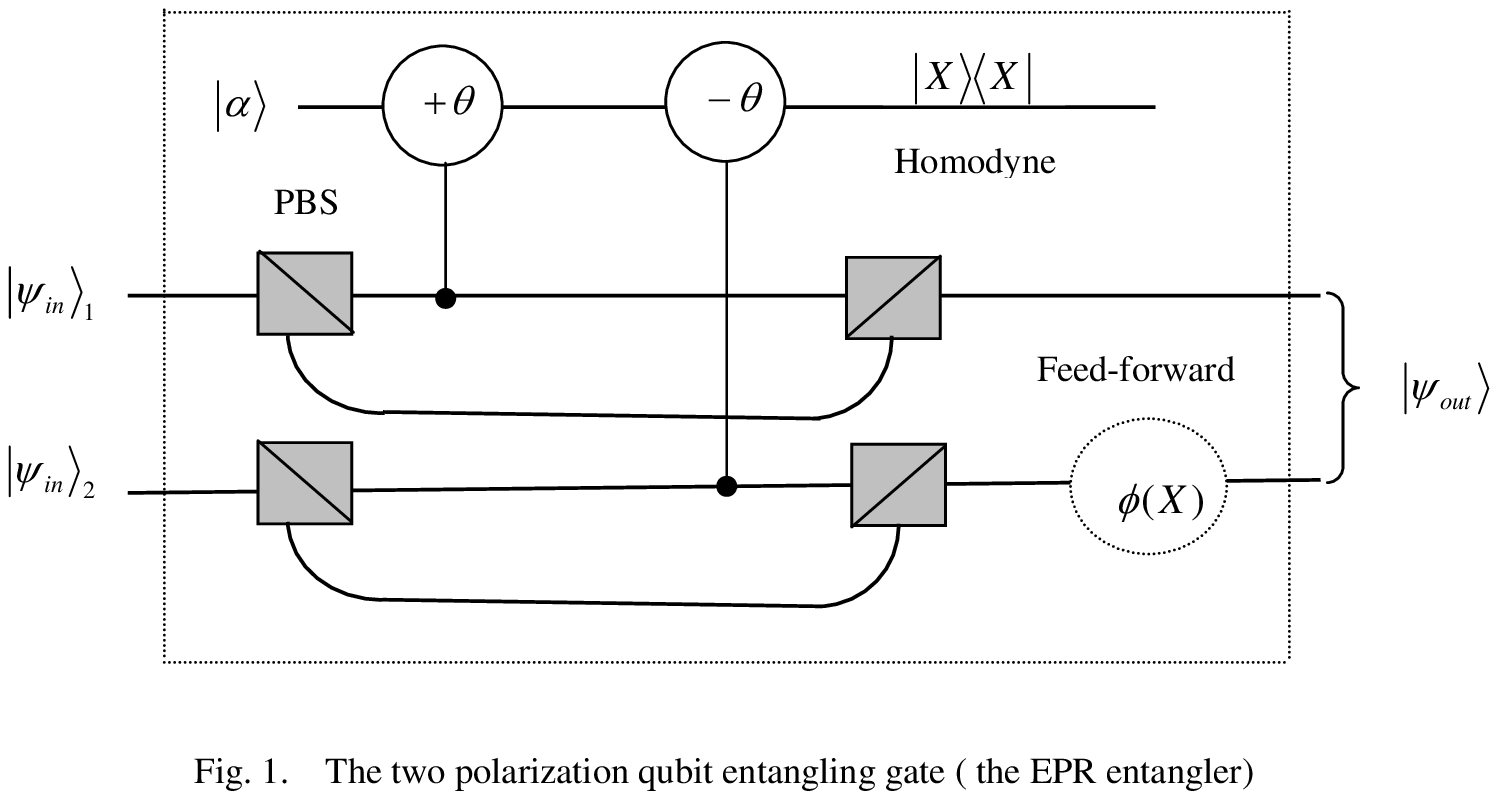}
\end {center}

   Consider two polarization qubits in signal modes initially prepared in the state
$|\psi\rangle_{1}=|\psi\rangle_{2}=\frac{1}{\sqrt{2}}(|H\rangle+|V\rangle)$.
These qubits are split individually on PBSs into spatial modes which
transmit $|H\rangle$ and reflect $|V\rangle$. Horizontal polarization
of each mode then interact with the intense coherence state in the
probe mode via cross-Kerr nonlinear medium. The action of the PBS¡¯s
and cross-Kerr nonlinearity will evolve the combined system
$|\psi\rangle_{1}|\psi\rangle_{2}|\alpha\rangle_{p}$ to
\begin{equation}\label{4}
|\psi\rangle_{T}=(|HH\rangle+|VV\rangle)|\alpha\rangle_{p}+(|HV\rangle|\alpha
\mathrm{e}^{\mathrm{i}\theta}\rangle_{p}+|VH\rangle|\alpha \mathrm{e}^{-\mathrm{i}\theta}\rangle_{p}),
\end{equation}
where we omit the normalization coefficients. Obviously, the states
$|HH\rangle$ and $|VV\rangle$ pick up no phase shift and remain
coherent with respect to each other, but the  states $|HV\rangle$
and $|VH\rangle$ pick up opposite phase shift $\theta$ which can be
distinguished by an X homodyne measurement. More specifically, with
$\alpha$ real, an X homodyne measurement conditions
$|\psi\rangle_{T}$ to
\begin{equation}\label{5}
|\psi_{x}\rangle_{T}=f(x,\alpha)[|HH\rangle+|VV\rangle]+f(x,\cos\theta)[|HV\rangle
\mathrm{e}^{\mathrm{i}\phi(x)}+|VH\rangle \mathrm{e}^{-\mathrm{i}\phi(x)}],
\end{equation}
where $f(x,\beta)=\exp[-\frac{1}{4}(x-2\beta)^{2}]/(2\pi)^{1/4}$ and
$\phi(x)=\alpha x\sin\theta-\alpha^{2}\sin2\theta$ (Mod $2\pi$).
$f(x,\alpha)$ and $f(x,\cos\theta)$ are two Gaussian curves with the
peaks located at $2\alpha, 2\alpha\cos\theta$ respectively. The
midpoint  and the distance between the two peaks are
$x_{0}=\alpha(1+\cos\theta)$ and $x_{d}=2\alpha(1-\cos\theta)$
respectively. Hence the singal states corresponding to different
measurement results $x$ are

\begin{eqnarray}
|\psi_{x}\rangle_{T}\sim \left\{
\begin {array}{ll}
|HH\rangle+|VV\rangle, &x>x_{0},\\
\mathrm{e}^{\mathrm{i}\phi(x)}|HV\rangle+\mathrm{e}^{-\mathrm{i}\phi(x)}|VH\rangle, &x<x_{0}.
\end{array}
\right.
\end{eqnarray}
Here we have chosen to call the even parity state
$(|HH\rangle,|VV\rangle)$ and the odd parity state
$(|HV\rangle,|VH\rangle)$. We have used the approximate symbol
($\sim$) in these equations as there is a small but finite
probability that the even parity state can occur for $x<x_{0}$. For
the odd parity state, a simple phase shift achieved via classical
feed forward allows it to be transformed to the even parity state.
Evidently, the action of this measurement method splits the even
parity terms nearly deterministically from the odd parity cases. So
we also call it the two-mode polarization non-demolition parity
detector, which acts as a near deterministic EPR entangler. This
gate is critical and forms the key element for  preparing
four-photon polarized-entangled states.

\section{Generation of four-photon polarized-entangled state }
  We now move our attention to the new protocol of generating the
four-photon polarized-entangled states described by Eqs.(1) and (2)
based on EPR entanglers. The setup schematic for preparing
$|\chi\rangle$ is illustrated in Fig.2.

\begin{center}
\scalebox{0.8}{\includegraphics* {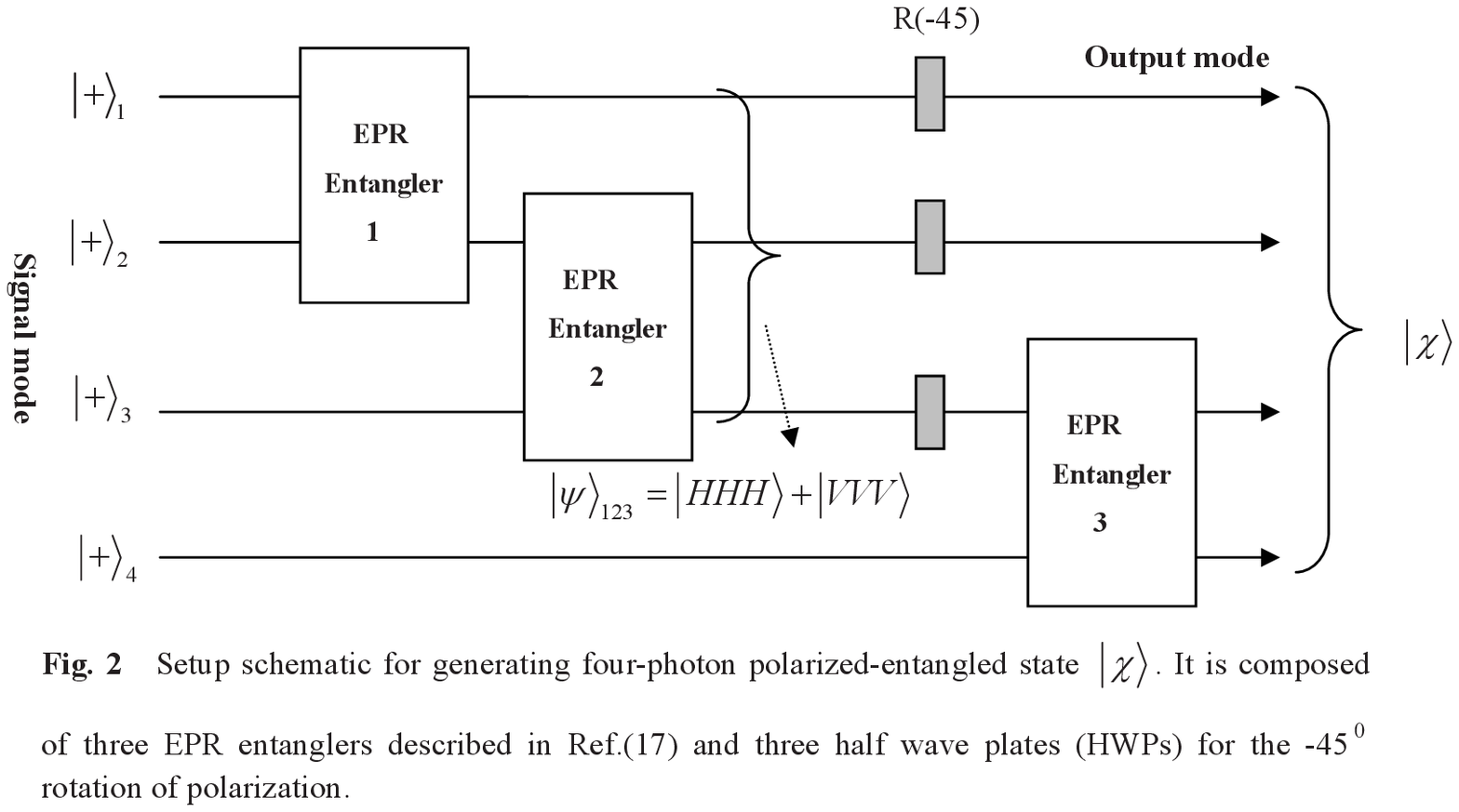}}
\end {center}

  The initial state of the system which consists of four photons in signal modes and a coherent probe beam is
$|\psi_{in}\rangle_{1234}=\otimes_{i=1}^{4}|+\rangle_{i}\otimes|\alpha\rangle$,
and $|+\rangle\equiv\frac{1}{\sqrt{2}}(|H\rangle+|V\rangle)$.
Firstly, the two photons in mode 1 and 2 pass through the EPR
entangler 1, then the two photons in mode 2 and 3 come into the EPR
entangler 2. The action of the two entanglers is to creat the
maximally entangled state $(|HHH\rangle+|VVV\rangle)$
or $(|HHV\rangle+|VVH\rangle)$, of course, the latter can be
transformed into the former by performing a bit flip on the third
photon qubit. In this step, we choose the three photons in mode 1, 2
and 3 to be in $(|HHH\rangle+|VVV\rangle)$. Subsequently, we exploit
three HWPs  for the $-45^{0}$ rotation of polarization R(-45) on mode
1, 2, and 3 respectively. The HWP is governed by the transformation
matrix
\begin{equation}
U_{\mathrm{HWP}}(\delta)=\left (\begin{array}{cc} \cos2\delta&-\sin2\delta\\
\sin2\delta&\cos2\delta
\end{array}\right),
\end{equation}
where $\delta$ is the rotation of polarization and the single-photon polarization basis states are denoted as
\begin{equation}
|H\rangle=\left (\begin{array}{c} 1\\
0
\end{array}\right),  ~~~|V\rangle=\left (\begin{array}{c} 0\\
1
\end{array}\right).
\end{equation}
At last, the two photons in mode 3 and 4 interact with the coherent
state in the EPR entangler 3. The whole system will evolve into
\begin{eqnarray}
|\psi'\rangle_{T}&=&(|HHHH\rangle+|VVHH\rangle+|HVVV\rangle+|VHVV\rangle)|\alpha\rangle \nonumber\\
                   &&{}+(|HHHV\rangle+|VVHV\rangle)|\mathrm{e}^{\mathrm{i}\theta}\alpha\rangle \nonumber\\
                  &&{}
                  +(|HVVH\rangle+|VHVH\rangle)|\mathrm{e}^{-\mathrm{i}\theta}\alpha\rangle.
\end{eqnarray}
 When performing the QND measurement in the entangler 3, we can
 obtain the $|\chi\rangle$ state or $\mathrm{e}^{\mathrm{i}\phi(x)}(|HHHV\rangle+|VVHV\rangle)
 +\mathrm{e}^{-\mathrm{i}\phi(x)}(|HVVH\rangle+|VHVH\rangle)$. For the latter, the phase factor $\mathrm{e}^{\pm\mathrm{ i}\phi(x)}$ could be
eliminated through a  dynamic phase shifter combined with classical
feed forward process. Thus we can combine the states
$(|HHHV\rangle+|VVHV\rangle)$  and $(|HVVH\rangle-|VHVH\rangle)$
together, the two combined states could be transformed to
four-photon polarized-entangled $|\chi\rangle$ state through
performing $\sigma_{x}$ operation on photon 4. Since the
$|\chi\rangle$ state and the $|cluster\rangle$ state is equivalent
under a local unitary transformation,  we can continue to perform
two HWPs for the $-45^{0}$ rotation of polarization R(-45) on mode 1,
and 2 respectively, and then we will obtain the cluster state. As a
matter of fact, the process of generating the cluster state can not
be so complex, it can be omitted some steps  if only we choose
another three-photon maximally entangled state when we make three
photons in mode 1, 2 and 3 entangled via the entangler 1 and 2. The
setup schematic is shown in Fig. 3.
\begin{center}
\scalebox{0.8}{\includegraphics* {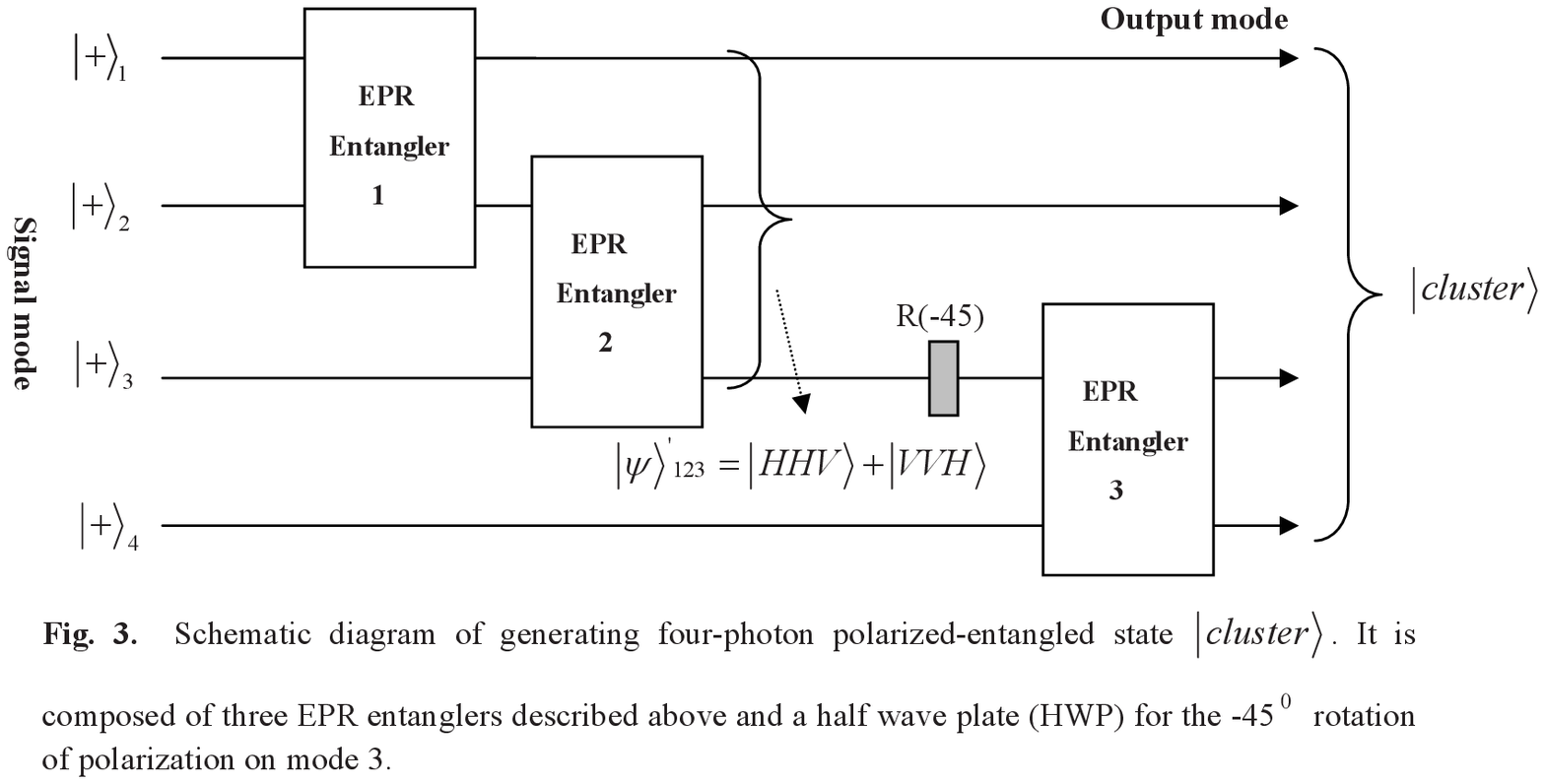}}
\end {center}

 As described above, at first the two photons in mode 1 and 2 interact in the entangler
 1, then the two photons in mode 2 and 3 pass through the entangler 2, the three photons are entangled. Here, we choose these photons to be in the maximally entangled
 state$(|HHV\rangle+|VVH\rangle)$. Afterward, we rotate the horizontal
 and vertical polarizations of the photon in mode 3 by utilizing
 R(-45), the whole system evolves into
\begin{equation}
|\psi'\rangle_{1234}=(|HHH\rangle+|HHV\rangle+|VVH\rangle-|VVV\rangle)_{123}\otimes|+\rangle_{4}.
\end{equation}
Next, the photons in mode 3 and 4 interact with cross-Kerr
nonlinearity in the entangler 3. The action of cross-Kerr
nonlinearity makes the four photons in
\begin{eqnarray}
|\psi''\rangle_{T}&=&(|HHHH\rangle+|HHVV\rangle+|VVHH\rangle-|VVVV\rangle)|\alpha\rangle \nonumber\\
                         &&{}+(|HHHV\rangle+|VVHV\rangle)|\mathrm{e}^{\mathrm{i}\theta}\alpha\rangle \nonumber\\
                         &&{}
                         +(|HHVH\rangle-|VVVH\rangle)|\mathrm{e}^{-\mathrm{i}\theta}\alpha\rangle.
\end{eqnarray}
It is obvious that the four-photon polarized state together with
coherent state $|\alpha\rangle$ is the cluster state. So we have
conditioned on an X homodyne measurement result either the state
$|cluster\rangle$ or
$\mathrm{e}^{\mathrm{i}\phi(x)}(|HHHV\rangle+|VVHV\rangle)+\mathrm{e}^{-\mathrm{i}\phi(x)}(|HHVH\rangle-|VVVH\rangle)$.
A simple phase shift achieved via classical feed forward and then  a
bit flip on the fourth polarization photon transforms the second
state   into the first.

\section{conclusion }
In summary, a simple experimental protocol is presented to generate
four-photon polarized-entangled states based on EPR entanglers. This
scheme has its distinct advantages: it uses only the basic tools in
quantum optical laboratories and can be implemented with the EPR
entanglers, which mainly involves a weak cross-Kerr nonlinearity
between signal modes and the probe coherent state followed by QND
measurement. This makes us  sure  it is feasible in the current
experimental technology. Moreover, it is possible to preparing the
four-photon state $|\chi\rangle$ or $|cluster\rangle$ with high
success probability since it primarily depends on three EPR
entanglers which can make pohtons entangled and detected near
deterministically. Thus, our scheme can be used  in QIP.

\vspace{0.5cm}

{\noindent\bf Acknowledgments}\\[0.2cm]

This work was supported by the National Natural Science Foundation
of China under Grant No: 10971247, Hebei Natural Science Foundation
of China under Grant No: A2012205013.


\end{document}